\documentclass[aps,pra, twocolumn, showpacs]{revtex4-2}
\usepackage{amsmath}
\usepackage{amsfonts}
\usepackage{amssymb}
\usepackage{fancyhdr}
\usepackage{color}
\usepackage[utf8]{inputenc}
\usepackage{mathtools}
\usepackage{graphicx}
\usepackage{float}
\usepackage{physics}
\usepackage{bbold}
\usepackage{url} 
\usepackage{color}
\usepackage{cancel}
\usepackage{array} 
\usepackage{latexsym}
\usepackage{mathrsfs}
\usepackage{enumitem}
\DeclareGraphicsExtensions{.pdf,.png,.jpg}
\usepackage{array}
\usepackage{verbatim}
\usepackage[title]{appendix}
\usepackage[colorlinks=true, urlcolor=blue,citecolor=blue,linkcolor=blue]{hyperref}

\begin{document}
\title{Simplified Quantum Optical Stokes observables and Bell's Theorem}     
\author{Konrad Schlichtholz, Bianka Woloncewicz, Marek \.Zukowski}
\affiliation{International Centre for Theory of Quantum Technologies (ICTQT),
University of Gdansk, 80-308 Gdansk, Poland} 
\begin{abstract}

We introduce a simplified form of Stokes operators for quantum optical fields that involve the known concept of binning. 
Behind polarization analyzer photon numbers (more generally intensities) are measured.
If the value obtained in one of the outputs, say H,  is greater the than in the other one, V, then the value of the simplified Stokes operator is, say, 1, otherwise it is -1. For equal photon numbers we put 0. 
Such observables do not have  all properties of the Stokes operators, but surprisingly can be employed  in Bell type measurements, involving polarization analyzers. 
They are especially handy for states of undefined number of photons, e.g. squeezed vacuum. 
We show that surprisingly they can lead to quite robust violations of associated Bell inequalities. 
\end{abstract}
\maketitle

\section{Introduction}

The discussion about what is the essence of quantumness,  started with first attempts of formulating of  quantum mechanics.
With the emblematic paper of Einstein, Podolsky and Rosen \cite{EPR}  the problem of completeness of quantum mechanics became a point of discussion among scientific community. This started with the response by Bohr  \cite{NONCLAS1}.
Many years later, after the paper of Bell \cite{BELL} the challenge of revealing non-classicality, in terms of violation of local realism,  has entered to the core of contemporary research.
All that in the meantime gained in importance with the emergence of quantum information and communication.  

The ultimate test of non-classicality  is violation of Bell inequalities. 
This is now also the essence of testing of device-independent quantum communication protocols.  
Formulations of Bell's theorem for situations of fixed numbers of particles have already a vast literature, and well established methods, see e.g. reviews \cite{Aspect2002,Brukner2012,Pan2012,Werner2001}. 
However, if one moves to situations with undefined numbers of particles, still the situation is quite open. This is of course e.g. the case of general
quantum optical fields. 
%A route to a definitive formulation is unknown. 
A lot of approaches are tested.

Polarization entanglement experiments are classic examples of experimental tests of Bell's inequalities.
The two photon experiments are a realization of two qubit-entanglement \cite{POLEXP1,POLEXP2}. 
A deceptively obvious  step in the direction towards optical fields of undefined photon numbers is to use quantum Stokes observables. 
The usual definition of these
runs as follows. If one assumes that the intensity of light is proportional to the photon number, then (standard) quantum Stokes observables are given by 
$\hat{\Theta}_i =
\hat{a^{\dagger}}_i\hat{a}_{i} -  \hat{a^{\dagger}}_{i\perp}\hat{a}_{i_{\perp}}, 
$
where  $\hat a$ is an annihilation operator.  Indices $i = 1,2,3$ mark three mutually unbiased (fully complementary) polarization analyzers settings.  The indexes, $i$ and   $i_{\perp}$ stand for two  orthogonal polarizations. 
E.g., one might choose the $i$'s to represent horizontal-vertical, $\{H,V\}$, diagonal-antidiagonal, $\{45^{\circ}, -45^{\circ}\} $, or right-left handed circular, $\{R,L\}$, polarization analyzer settings. 
The zeroth Stokes operator is given by the total photon number  operator $\hat \Theta_0 = \hat{N}= \hat{a^{\dagger}}_i\hat{a}_{i} +  \hat{a^{\dagger}}_{i\perp}\hat{a}_{i_{\perp}}$ \cite{PHOTONS}. 

If we are interested in the degree of polarisation of  light  we use
$\big(\frac{\sum_i\expval{\Theta_i}^2}{\expval{\Theta_0}^2}\big)^{1/2}$. 
Obviously,  this parameter is not a formal quantum observable (a selfadjoint linear operator). Neither is $\frac{\expval{\Theta_i}}{\expval{\Theta_0}}$. 
This  is  one of the reasons why attempts to build Bell inequalities using such parameters and their correlators for observation stations $A$ and $B$ in the form of $\frac{\expval{\Theta^A_i\Theta^B_j}}{\expval{\Theta^A_0\Theta^B_0}}$ fail and lead to misleading conclusions \cite{SINGLE1}. This is because such attempts involve additional assumptions, beyond the usual ones for Bell inequalities, which limit the range of local hidden variable theories for with such Bell inequalities musts hold.  
%require additional assumptions.%,
%except the ones on which Bell inequalities rest.

Bell inequalities for Stokes parameters can be formulated if one introduces normalized Stokes observables
\cite{Stokes:original}, \cite{ZUKUPRA},\cite{ZUKUBELL}: \begin{equation}
 \hat S_{j}  =  \hat \Pi \frac{\hat n_{j} - \hat n_{{j}_\perp}}{\hat n_{j} + \hat n_{{j}_\perp}} \hat \Pi,
\label{STOKES0}
\end{equation}
where $\hat \Pi=\mathbb{I}-\ket{\Omega}\bra{\Omega}$, and 
$\ket{\Omega}$ is the vacuum state (of the optical beam in question). %and $\hat n_j$ are number operators for which  $j=1,2,3$ and $j_{\perp}$ refer to vectors from  different polarization  basis as it was described  in 
%the case of traditional Stokes operators.
It has been shown that such operators allow construction of stronger entanglement criteria and they are a handy tool for formulation of Bell inequalities. One of their properties, which is crucial in this case, is the fact that, such operators have spectrum which consists on all rational numbers from $-1$ to $1$. That is, they have the basic property of observables which allows to derive the CHSH-Bell inequalities. Thus a derivation of a version of CHSH inequality applicable for such Stokes operators essentially is a replacement procedure. With the  recent development  of  measurement techniques allowing photon number resolving detection \cite{WALMSLEY},\cite{WEAKHOMO-OPTICAL-COHERENCE} the discussion about normalized Stokes parameters stops to be only theoretical and its use in experiments is becoming feasible.

Note that what makes Pauli operators so straightforwardly applicable  for Bell inequalities is their dychotomic nature.  One of the attempts to construct  field operators of a similar property was the  formulation of pseudo-spin operators. E.g. the $z$ component of pseudo spin is $(-1)^{\hat n}$, where $\hat n$ is total photon number in the given optical mode \cite{Pspin1},\cite{Pspin2}.
The spectrum of  pseudo-spin operators is the same as spectrum of Pauli matrices, but their use introduces great difficulties from experimental point of view.
Even a loss of one photon (due to e.g. detector inefficiency) or  a single dark count reverses the result of a  measurement. 
%That  leads to false interpretation of the observed phenomena.

We analyze at a simpler approach, which  leads to proper Bell inequalities for polarization measurements of quantum optical  fields.
Our aim is  to construct a family of operators that would have usual spectrum for Bell experiments,  and  is robust with respect to experimental noise. 
We present polarization quantum field observables that have spectrum limited to $\pm1$ and $0$. Our initial ideas on such binning can be found in \cite{Thesis}. The approach to binning presented here is concurrent of the method used in \cite{PhysRevA.104.043323} in the context of correlation in Bose-Einstein condensates.
With  the observables we construct Bell inequalities. We test their resilience under losses and noise for $2 \times 2$ mode bright squeezed vacuum and bright GHZ radiation. 
The  observables are realizable in the laboratory with  standard measurement devices. They are described in the next section.

\section{New operators: sign Stokes operators}
It was shown that Bell inequalities constructed with normalized Stokes operators can be violated by macroscopic states of light like $2 \times 2$ Squeezed vacuum (BSV) and its  GHZ-like generalization (BGHZ) \cite{ZUKUBELL}, \cite{BGHZ}. However, for higher mean number of photons violation of Bell inequalities by the states is quickly damped. This results in lowering of the threshold values for pumping strength after which violation cannot be observed. 

We  address those problems by another normalization scheme, see \cite{PhysRevA.104.043323}, based on the so-called binning, which we call Sign approach normalization. To obtain new operators we use the  sign function and apply it to Stokes operators:  

\begin{align}
\begin{split}
\hat G_i&=\mathrm{sign}(\hat n_i-\hat n_{i_\bot})\\
&= \mathrm{sign}(\hat \Theta_i)=\mathrm{sign}(\hat U_i(\hat n_H-\hat n_{V})U_i^{\dagger}),
\label{sign}
\end{split}
\end{align}
where indices $H$ and $V$ refer to horizontal and vertical polarizations and operator $\hat U_i$ is an unitary transformation which transforms polarization modes $H$, $V$ into another orthogonal pair of in general elliptic polarization modes $i$-th and $i_\perp$ (for examples see: Appendix \ref{exper}). 
From (\ref{sign}) we see that the eigenstates of the $i$-th operator are photon number states of a given mode of a given polarization i.e. $\ket{j}_i$ and
$\ket{k}_{i_{\perp}}$. 
These kets symbolically represent photon number states: $\ket{j}_i
\ket{k}_{i_{\perp}}=\frac{1}{\sqrt{j!k!}}\hat{a}_i^{\dagger j}\hat{a}_{i_\perp}^{\dagger k}\ket{\Omega}.$
The  spectral form  of (\ref{sign}) is given by:
\begin{align}
\begin{split}
\hat G(i)=\sum_{k>j}\Big(&\ket{k}_i\ket{j}_{i_\bot}\bra{k}_i\bra{j}_{i_\bot}\\
&-\ket{j}_i\ket{k}_{i_\bot}\bra{j}_i\bra{k}_{i_\bot}\Big).
\label{SPEC}
\end{split}
\end{align}
Formula (\ref{SPEC})  clearly shows that new  operators are well-defined Hermitian ones and that $ \hat G_i $ have three eigenvalues $\pm1$ and $0$. 

Action of $\mathrm{sign}$ function  on Stokes operators can be regarded as some form of the binning strategy used in the context of polarization measurements. Binning strategies are e.g.
used in homodyne schemes for observing non-classicality \cite{bin1,bin2,bin3,bin4}. 

We shall call the new operators  sign Stokes operators.
We denote by $\hat G_1$ the sign operator the eigenstates of which refer to $\{D,A\}$ polarisation basis, and  by $\hat G_2$ and  $\hat G_3$ for respectively $\{R,L\}$ and $\{H,V\}$ bases.

From all that was  said above, one can easily conclude that
sign  operators
share some  properties of Stokes and normalized Stokes operators. Also, once one has a photon-number resolving detection setup, the data collected in each run  allows to compute the obtained values of each of  Stokes operators for the given basis $i$: standard, normalized, and sign ones, as they depend solely on the measured $n_i$ and $n_{i{\perp}}$. 
Thus, as we see, the new approach is in fact just a new form of data analysis.  
Further, in order to measure different sign operators $\hat G_i$, i.e. in order to move from $i$ to $i'$, it is enough to change the polarization analysis  basis (see Appendix \ref{exper}). 
However, not all properties of quantum Stokes and normalized Stokes operators are shared by quantum sign Stokes operators.

\subsection{Stokes vector formed  out of sign Stokes operators} 

 For the standard Stokes operators one can construct a Stokes vector i.e. $\langle \hat{\overrightarrow{\Theta}}\rangle_\psi=(\langle \hat \Theta_1\rangle_\psi,\langle \hat \Theta_2\rangle_\psi,\langle \hat \Theta_3\rangle_\psi)$ where $\ket{\psi}$ us an arbitrary  state of the optical field              The norm of this vector fulfills : $||\langle \overrightarrow{\Theta}\rangle_\psi||\leq
\expval{\hat \Theta_0}_{\psi}$. 

We can construct an analogue vector for normalized Stokes operators and $||\langle \hat{\vec{S}}\rangle_\psi||\leq
\expval{\hat S_0}_{\psi}\leq 1$ \cite{ZUKUPRA}. 
These  norms remain invariant  under unitary transformation $\hat U$ between mutually unbiased polarisation basis i.e.:
$||\langle \hat {\overrightarrow{\Theta}}\rangle_\psi||=||\langle \hat{\overrightarrow{\Theta}}\rangle_{\psi'}||$
where $\ket{\psi'} = \hat U^\dagger\ket{\psi}$
and  
$||\langle \hat {\overrightarrow{S}}\rangle_\psi||=||\langle \hat{\overrightarrow{S}}\rangle_{\psi'}||$.

Thus norm of Stokes vectors, standard and normalized, is  constant under unitary   of the triads polarization analysis bases . 
These features of Stokes  play a key role  in construction of  entanglement indicators involving Stokes operators. 

Such properties are not shared by sign operators. Let us construct a vector like triad
$\langle \hat{ \overrightarrow{G}}\rangle_\psi =
(\langle \hat G_1\rangle_\psi,\langle \hat G_2\rangle_\psi,\langle \hat G_3\rangle_\psi)$. It can be shown that $||\langle \overrightarrow{G}\rangle_\psi|| \neq ||\langle \overrightarrow{G}\rangle_{\psi'}||$. 
It is enough to find one counterexample. Consider state $\ket{\psi}   = \ket{3_H,0_V}$ i.e. Fock state with $3$ photons polarised horizontally. It can be easily checked that for this state $||\langle \overrightarrow{G}\rangle_{\psi}||=1$.
After performing a $\mathit{SO}(2)$ rotation of
$\ket{\psi}$  by  $\pi/8$ we get
 $||\langle \overrightarrow{G}\rangle_{\psi'}|| \approx 1,5$ Thus, 
$||\langle \overrightarrow{G}\rangle_{\psi} || \neq  ||\langle\overrightarrow{G}\rangle_{\psi'}||
$, i.e. the norm is not an invariant of of the unitary transformations 
and additionally it is not bounded by $\langle \hat G_0\rangle$. This fact prohibits one to use methods of construction of entanglement indicators presented in \cite{ZUKUPRA}, which works via a simple replacement of Pauli operators in entanglement conditions for qubits, by Stokes operators, standard or normalized. Still, as we shall see there is no obstacle to use this method in the case of construction of Bell inequalities.

Rotational {\em covariance} of polarization variables is not a necessary feature required to derive Bell inequalities (also, see \cite{RotInv} for consequence of demanding exactly that). This allows one to construct CHSH and CH inequalities for fields with sign Stokes observables.
The main goal of proposing new operators, as we have said earlier, is to enable detecting nonclassicality for optical fields of higher mean intensity than in case of normalized Stokes operators. Also questions such as resistance to noise and losses of new operators are  of our concern.

%In the following we will consider firstly CHSH inequality and secondly CH inequality.

\subsection{CHSH inequality}\label{sectionchsh}
We start with defining local hidden values which predetermine output of a measurement of sign Stokes operators (\ref{sign}). 
We denote the local hidden variables by $\lambda$. The functions 
$I^X(i,\lambda)$ and $I^X(i_{\perp},\lambda)$
give the predetermined outcomes of intensity measurements of  polarizations $i$, $i_\perp$ in the local beam for observer $X$.
The measurement is done with tuneable setting in presence of $\lambda$.
We define local hidden values for sign operators as $G^X(i,\lambda)=\mathrm{sign}(I^X(i,\lambda)-I^X(i_\perp,\lambda))$. 
These local hidden values are $\pm1$ and $0$, thus one can use standard methods to derive CHSH inequality.
We assign the following set of settings $i = \theta,\theta'$ for first observer and $i= \phi, \phi'$ for second observer. 
Resulting CHSH inequality reads:  
\begin{align}
\begin{split} 
&|\langle G^{1}(\theta,\lambda) G^{2}(\phi,\lambda)+ G^{1}(\theta,\lambda) G^{2}(\phi',\lambda)\\
&+ G^{1}(\theta',\lambda) G^{2}(\phi,\lambda)- G^{1}(\theta',\lambda) G^{2}(\phi',\lambda)\rangle_{LHV}|\leq 2, \label{bsvin1}
\end{split}
\end{align}
%However, this inequality cannot be violated by states with a significant vacuum component e,g. such is the case of our working example, see next Sections,  the  (polarization) four-mode squeezed vacuum state. 
However, this inequality cannot be violated by states with a significant vacuum component, e,g. the (polarization) four mode  squeezed vacuum state, which will be our working example, see next Sections. 
This situation is analogous to the case of normalized Stokes operators, see \cite{ZUKUBELL}. 
Following ideas of \cite{ZUKUBELL} we modify sign Stokes operators as follows:
\begin{equation}
\hat G^X(i)\rightarrow \hat G^{X-}(i)=\hat G^X(i)-\ket{\Omega^X}\bra{\Omega^X},
\end{equation}
what allows for reduction of  the impact of vacuum term which often appears with  the highest probability. 
Also local hidden values need to be modified:
\begin{itemize}
	\item{$G^{X-}(i,\lambda)=\mathrm{sign}(I^X(i,\lambda)-I^X(i_\perp,\lambda))$ if \newline $I^X(i,\lambda)+I^X(i_\perp,\lambda)\neq 0$
}
  \item{$G^{X-}(i,\lambda)=-1$ if $I^X(i,\lambda)+I^X(i_\perp,\lambda)=0$} 
\end{itemize}

As this modification does not change local hidden values $G^{X-}(i,\lambda) \in \{0, \pm1\}$ we use the following CHSH inequality: 

\begin{align} 
\begin{split}
&|\langle G^{1-}(\theta,\lambda) G^{2-}(\phi,\lambda)+ G^{1-}(\theta,\lambda) G^{2-}(\phi',\lambda)\\
&+G^{1-}(\theta',\lambda) G^{2-}(\phi,\lambda)- G^{1-}(\theta',\lambda) G^{2-}(\phi',\lambda)\rangle_{LHV}|\leq 2. \label{bsvin}
\end{split}
\end{align}

\begin{figure}[h!]
%\centering
\includegraphics[width=\textwidth/2]{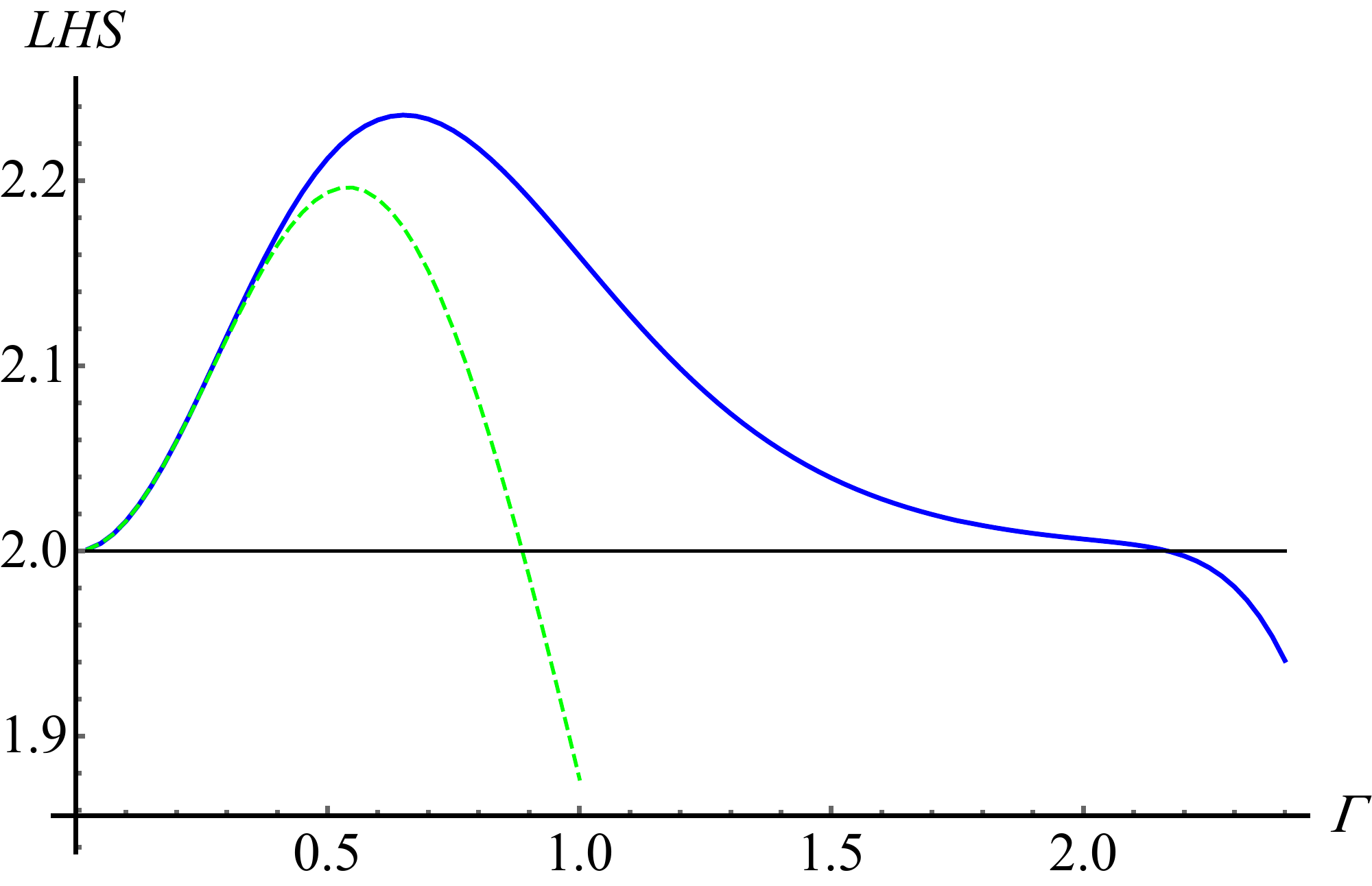} 
\caption{$LHS$ of  CHSH inequality based on sign operators (\ref{bsvin}) - blue curve, and CHSH inequality based on normalized Stokes operators \cite{ZUKUBELL} -green dashed curve- in a function of amplification gain $\Gamma$ for $BSV$ state. 
We perform numerical cutoff on terms with more that 150 photons. 
The threshold values of amplification gain ($\Gamma_{tr}$), such that for all  $\Gamma < \Gamma_{tr}$ CHSH inequalities are violated, are $\Gamma_{tr} \approx 0.88$ for normalized Stokes operators \cite{ZUKUBELL} and  $\Gamma_{tr} \approx 2.16$ for sign Stokes operators. 
Thus, with sign Stokes operators  the range of violation with respect to amplification gain is much larger that in case of normalized Stokes operators.}
\label{bsv}
\end{figure}

\subsubsection{Violation of Bell inequality for four mode squeezed vacuum - asymptotic behaviour}
\label{asym}

We are going to analyze how the use of sign Stokes operators in CHSH inequality helps to reveal  the nonclassicality of quantum states. Our working example is  $2\times 2$ mode squeezed vacuum state (BSV) which is the generalization of EPR singlet.  
It  reads
\begin{equation}
\ket{\psi_-}=\frac{1}{\cosh^2(\Gamma)}\sum_{n=0}^\infty\sqrt{n+1}\tanh^n(\Gamma)\ket{\psi^n},
\end{equation}
where  $\Gamma$ is amplification gain,  and
\begin{align}
\begin{split}
\ket{\psi^n}=&\frac{1}{\sqrt{n+1}}\\
&\sum_{m=0}^n(-1)^m\ket{n-m}_{H_1}\ket{m}_{V_1}\ket{m}_{H_2}\ket{n-m}_{V_2}.
\end{split}
\end{align}
Subscripts $q=1,2$ in $H_q$ and $V_q$ specify to which of the two beams particular mode corresponds. Amplification gain determines the  intensity of pumping field and thus $\Gamma$ sets expectation value of intensity of the BSV state.

Let  $\theta$, $\theta'$, $\phi$ and $\phi'$ be the settings used in CHSH inequality. These settings describe the angles by which the measurement polarization basis is rotated in relation to $\{H,V\}$ basis by $\mathit{SO}(2)$ rotation. 
%Let us consider a case  with $\theta$, $\theta'$, $\phi$ and $\phi'$ describing  the  angles (setting) by which   measurement polarization basis is rotated in relation to $\{H,V\}$ basis by $\mathit{SO}(2)$ rotation. 
We chose: $\theta=0$, $\theta'=\pi/4$, $\phi=\pi/8$ and  $\phi'=-\pi/8$.  
It was shown that these settings are  optimal in case of violation of CHSH inequality with normalized Stokes operators for BSV \cite{ZUKUBELL}.

%FIG. \ref{bsv} shows quantum predictions of LHS of  (\ref{bsvin}) for BSV state ($\bra{\psi_-}CHSH\ket{\psi_-}$) in a function of amplification gain $\Gamma$. Also results for analogous inequality for normalized Stokes operators reported in \cite{ZUKUBELL} are presented on FIG. \ref{bsv}.

FIG. \ref{bsv} shows quantum predictions of LHS of CHSH inequality for sign Stokes operators (\ref{bsvin})  LHS of CHSH for normalized Stokes parameters for BSV taken from from \cite{ZUKUBELL}  in a function of amplification gain $\Gamma$. From now on  we denote  LHS of (\ref{bsvin}) for BSV by $\bra{\psi_-}CHSH\ket{\psi_-}$. 
Sign Stokes operators allow for violation of CHSH for wider range of amplification gain that is  up to $\Gamma_{tr}\approx 2.16$.
For normalized Stokes operators this value is  significantly lower i.e. $\Gamma_{tr}=0.8866$. 
Thus with sign Stokes  operators it is possible to reveal nonclassicality of  BSV with higher value of expectation value of intensity (brighter light).

%This shows that sign Stokes operator allows for witnessing nonclasicality for states with higher expectation values of intensity than in case of normalized Stokes operators$-

From the other hand in FIG. \ref{bsv}  we can see that for $\Gamma\approx 2.1$  LHS for sign Stokes operators drops down suspiciously suddenly.
%starts to decrease faster and no longer violate inequality . 
We presume that such a behaviour might be a consequence of a cutoff  performed on BSV state (the superposition of  $\ket{\psi^n}$s is considered up to $n=150$ photons). Such a statement requires further investigation.

However, before we pursue further we need to make a remark
about the BSV state  and sign and Stokes operators. First, BSV is invariant under any unitary operation performed on both observers.
Thus, the measured expectation value depends only on the difference $\theta-\phi$. Also, the components of BSV, states $\ket{\psi^n}$ are  orthogonal: $\langle{\psi^n|\psi^k}\rangle = 0$ for $n \neq k$, 

Note that standard, normalized and sign Stokes operators  are composed from functions of photon number operators, that do not change number of photons. Thus, LHS of CHSH inequality  consists of two terms:  {vacuum term}, that is CHSH inequality averaged over vacuum component of BSV and {non-vacuum term}.  
The vacuum term can be easily calculated:
\begin{equation}
\sum_{j= \theta,\theta'}\sum_{k=\phi,\phi'}\expval{\hat G^{1-}(j) \hat G^{2-}(k)}{\Omega} = \frac{2}{\cosh^4\Gamma}.
\label{VACUUM}
\end{equation}

The non-vacuum terms ($LHS_{nv}$)  results from the expectation values of $\ket{\psi^n}$.  Note that as $\Gamma$ increases the role of non-vacuum terms in LHS of  CHSH inequality increases too. For small $\Gamma$ the  contribution of vacuum term  dominant. 

\begin{figure}[h!]
\centering
\includegraphics[width=0.49\textwidth]{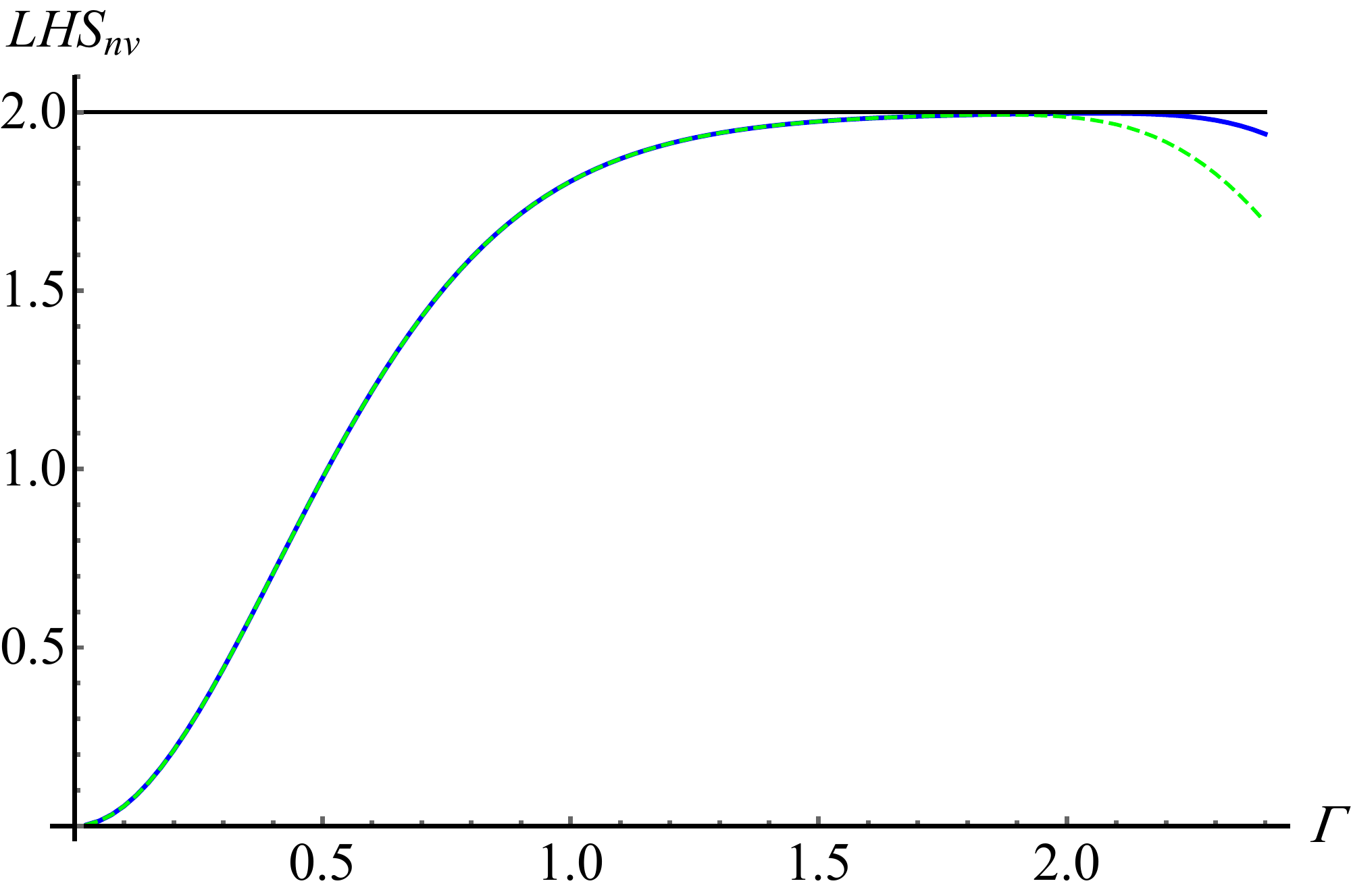} 
\caption{Non-vacuum term of $LHS$ of the inequality (\ref{bsvin}) as a function of amplification gain  $\Gamma$ for the BSV state. A blue curve represents calculations with 150 photons and a green dashed curve with 100 photons. A change of character of the function from increasing to decreasing starts in case of 100 photons for smaller $\Gamma$ than for 150 photons. This suggest that this qualitative change in function behaviour is only an artifact of including not sufficient amount of terms.}
\label{bsv2}
\end{figure}

In FIG. \ref{bsv2} non-vacuum contribution to LHS of (\ref{bsvin}) is presented. The calculation is performed for BSV state  truncated up to  $n=150$ - blue curve and $n=100$ - green curve. Both curves  asymptotically go to 2 (classical bound) up to some point for which they both start to decrease. Note that the curve for $n=100$ decreases faster than the curve for $n=150$.  it is highly probable that the decrease is conditioned by not including components with high enough number of photons and   the non-vacuum term of  LHS of $\ref{bsvin}$ goes asymptotically to 2 from the left. The vacuum term goes asymptotically to 0 from the right, see (\ref{VACUUM}). Thus, our hypothesis is that  CHSH inequality with  sign Stokes operators is violated for BSV for any $\Gamma$. Below we present a reasoning  based on numerical calculation.

%The blue curve stands for calculations performed up to components of the state with 150 photons. The green dashed curve shows the case for up to  components with 100 photons. 
%Both curves  in FIG. \ref{bsv2} asymptotically go to 2 (classical bound) up to some point for which they start to decrease rapidly.
%It is  highly probable that the reason of such a behaviour is that, due to computational limitations  in our calculation we do not consider terms with high enough number of photons.
%Our hypothesis is that  
%in LHS of $\ref{bsvin}$ we have a term that goes asymptotically to 2 from  the left and a term which go asymptotically to 0 from the  right (vacuum term, see (\ref{VACUUM})).
Let us analyze expectation values (\ref{bsvin}) for states $\ket{\psi^n}$ i.e. $\langle CHSH\rangle_{\psi^n}$ and compare them with LHS of the analogue  expression for normalized Stokes operators from \cite{ZUKUBELL}.
%The relation between LHS of $\ref{bsvin}$ and the amplification gain requires further investigation. Let us  analyze expectation values of inequality (\ref{bsvin}) for states $\ket{\psi^n}$ which we will denote as $\langle CHSH\rangle_{\psi^n}$.

\begin{figure}[h!]
\centering
\includegraphics[width=0.49\textwidth]{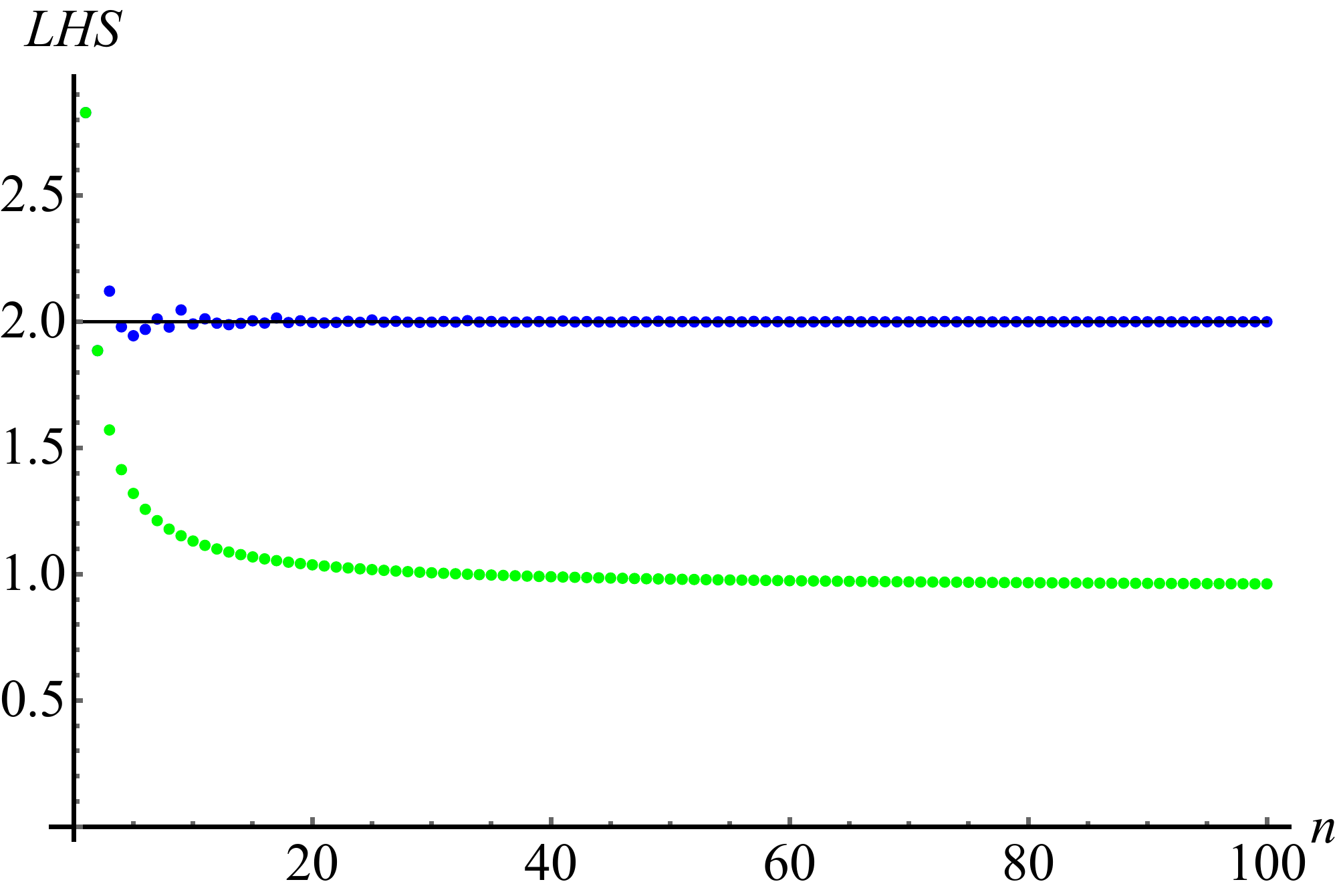} 
\caption{  $LHS$ of CHSH inequality based on sign (blue points) and  normalized (green points) Stokes operators for $\ket{\psi^n_-}$ as a function of $n$.
The  points: $n=1$  and $n=2$ of both approaches coincide. For normalized Stokes operators  only singlet state contribute to the violation and all points  converge to $1$ which is the bound for separable states. In case of sign Stokes operators  all points are concentrated around  the classical bound for CHSH inequality. Still, not for all $n$'s violation of classical bound occurs (for details see Appendix \ref{psinV}).  
%\mz{1. for n=2 definitely blue and green coincide. This must be marked. 2. Are you sure that for some of the component states the blue values are BELOW 2? If so this requires an explanation, e.g. for the simplest example}
}\label{psi}
\end{figure}

FIG.\ref{psi} shows results for $n=1,...,100$  for sign Stokes operators and normalized Stokes operators. For normalized Stokes operators only for $n=1$  we get $LHS \geq 2$. For sign operators values of $\langle CHSH\rangle_{\psi^n}$ concentrate around $2$ with growing $n$. More detailed analysis, see Appendix \ref{psinV}, reveals two patters: an oscillating one for odd $n$'s and a pattern converging to $2$ from bellow for even $n$'s. 
The period of odd $n$'s is equal to $T=8$ in the sense that  points  $n=2k+1$ and  $n=2k+1+T$ where $k \in \mathbb{Z}$  correspond to e.g. two adjacent maximums in the pattern.
%Note that only calculating LHS  for the sum over  $T$  or $pT$ states $\ket{\psi^n}$ corresponding to the entire period i.e. from $n =2k+1$ to $n=2k+1+T$  ensures the violation of (\ref{bsvin}). Taking any set of  $\ket{\psi^n}$ that is not bounded by  points $n =2k+1$ to $n=2k+1+T$ with period , where $p$ is an integer, does not ensure violation of CHSH. 
The even pattern also has internal structure repeatable with $T=8$. 

\begin{figure}[h!]
\centering
\includegraphics[width=0.49\textwidth]{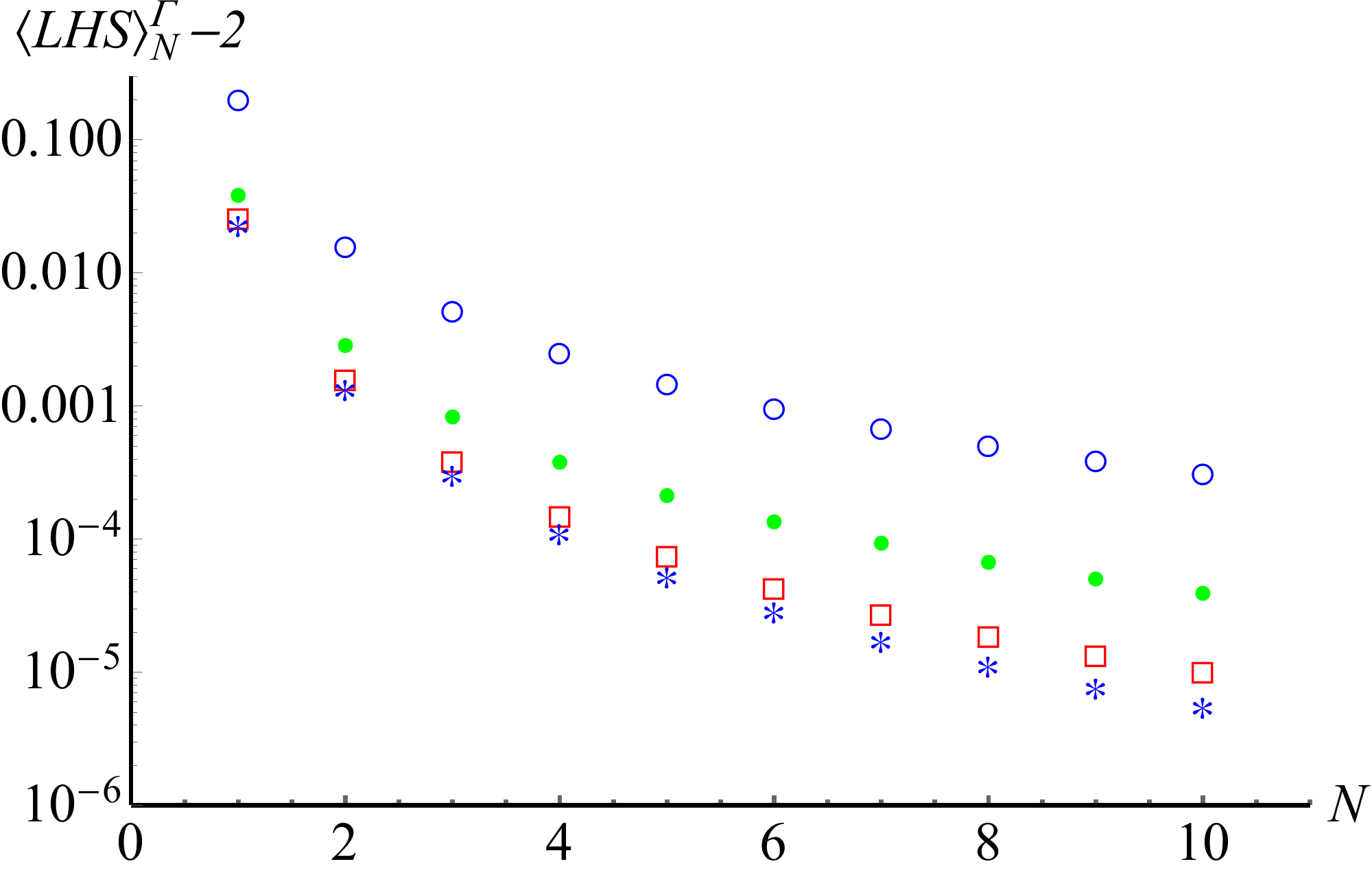} 
\caption{$\langle LHS \rangle_N^\Gamma$ versus $N$. $\Gamma=1$ circles, $\Gamma=2$ dots, $\Gamma=3$ squares, $\Gamma\rightarrow\infty$ stars. For any $\Gamma$ values of $\langle LHS \rangle_N^\Gamma$ go to $2$ with growing $N$. The case of $\Gamma\rightarrow\infty$ bounds $\langle LHS \rangle_N^\Gamma$ from bellow.}
\label{blok}
\end{figure}

Periodicity of the pattern provide us the natural grouping of $\ket{\psi^n}$ for the regarded problem. Let us examine a weighted average of 
$\langle LHS \rangle_N^{(\Gamma)}$ for a given $\Gamma$ over $N$-th period for $\Gamma>2$ with weights $w_n$:
\begin{multline}
\langle LHS \rangle_N^{(\Gamma)}=\\
\frac{\sum_{n=1+T (N-1)}^{T+T (N-1)}\langle CHSH\rangle_{\psi^n}|\braket{\psi_-^n}{\psi_-}|^2}{\sum_{n=1+T (N-1)}^{T+T (N-1)}|\braket{\psi_-^n}{\psi_-}|^2},
\end{multline}
where $|\braket{\psi_-^n}{\psi_-}|^2=(n+1)\frac{\tanh^{2n}{\Gamma}}{\cosh^4{\Gamma}}$. Fig. \ref{blok} shows values of $\langle LHS \rangle_N^{(\Gamma)}-2$ for $\Gamma=1,2,3$ and $\Gamma\rightarrow\infty$. 

We observe that $\langle LHS \rangle_N^{(\Gamma)}$ is a decreasing function of $\Gamma$. All calculated values of $\langle LHS \rangle_N^{(\Gamma)}$ exceed $2$ and violate the inequality (\ref{bsvin}). 
%Thus, every period is dominated by the violating points of the odd pattern. 
Also $\langle LHS \rangle_N^{(\Gamma)}$ for any given $\Gamma$ converges to $2$ with growing $N$.  Moreover the curve corresponding to $\Gamma\rightarrow\infty$ is the most relevant for our analysis  because it  bounds the  $\langle LHS \rangle_N^{(\Gamma)}$ from bellow. 

Let us observe that non-vacuum term $LHS_{n{v}}$ can be written as the weighted average of $\langle LHS \rangle_N^{(\Gamma)}$:
\begin{multline}
LHS_{nv}=
\sum_{N=1}^\infty \langle LHS \rangle_N^{(\Gamma)} \sum_{n=1+T (N-1)}^{T+T (N-1)}|\braket{\psi_-^n}{\psi_-}|^2.
\end{multline}

Assuming that there is no change in pattern of  $\langle CHSH\rangle_{\psi^n}$ as $n$  increases (see Appendix \ref{psinV} for the argumentation) we can bound from bellow the value of $LHS_{nv}$ by replacing $\langle LHS \rangle_N^{(\Gamma)}$ with $2$:
\begin{align}
\begin{split}
LHS_{nv}&\geq\sum_{n=1}^\infty 2(n+1)\frac{\tanh^{2n}{\Gamma}}{\cosh^4{\Gamma}}\\
&=2(\tanh^2{\Gamma}+\sech^2{\Gamma}\tanh^2{\Gamma}).\label{asnv}
\end{split}
\end{align}
Equality due to our assumptions should be only reached in the limit of $\Gamma\rightarrow\infty$. This is because in the regime of high values of $\Gamma$ only terms with high $n$ are significant and $\langle LHS \rangle_N^{(\Gamma)}$ from the assumption reach $2$ only when $N\rightarrow\infty$. Expression (\ref{asnv}) as expected has an asymptotic value 2. If we add the vacuum term (\ref{VACUUM}) to the RHS of (\ref{asnv}) we obtain constant function $2$. Thus, finally we obtain:
\begin{equation}
    \bra{\psi_-}CHSH\ket{\psi_-}\geq 2.
\end{equation}
This strongly suggest that violation of (\ref{bsvin}) with $\ket{\psi_-}$ is possible for any finite $\Gamma$ and numerically obtained existence of the threshold value of $\Gamma$ is due to computational limitations.


%\begin{figure}[h!]
%\centering 
%\includegraphics[width=\textwidth/2]{diff.pdf} 
%\caption{Difference between first 10 terms of the non-vacuum term of $LHS$ of the inequality (\ref{bsvin}) and 10 first terms of approximation (\ref{asnv}) in function of parameter  $\Gamma$ for the BSV state. The difference is positive expect for $\Gamma=0$ and in the limit of $\Gamma\rightarrow\infty$ goes to 0. }
%\label{diff}
%\end{figure}
\subsection{CHSH inequality with losses} \label{losses}
One of the critical aspects of experimental realization of Bell experiments are detectors with high efficiency $\eta$ of detection. Here we will analyze the critical value of efficiency $\eta_{c}$ such that for $\eta<\eta_{c}$ one can not observe violation of (\ref{bsvin}). We model inefficient detectors following \cite{ZUKUBELL}. Such detector can be described as perfect detector ($\eta=1$) in front of which there is beamspliter with transitivity  $\sqrt{\eta}$.  We denote number of photons which reach detectors as $k$. From those photons only $\kappa\leq k$ counts are registered due to losses on beamspliter. For such case probability of measuring $\kappa$ photons is given by the binomial distribution:
\begin{equation}
p(\kappa|k)= {k\choose \kappa}\eta^{\kappa}(1-\eta)^{k-\kappa}.
\end{equation}

\begin{figure}[h!]
\centering
\includegraphics[width=0.49\textwidth]{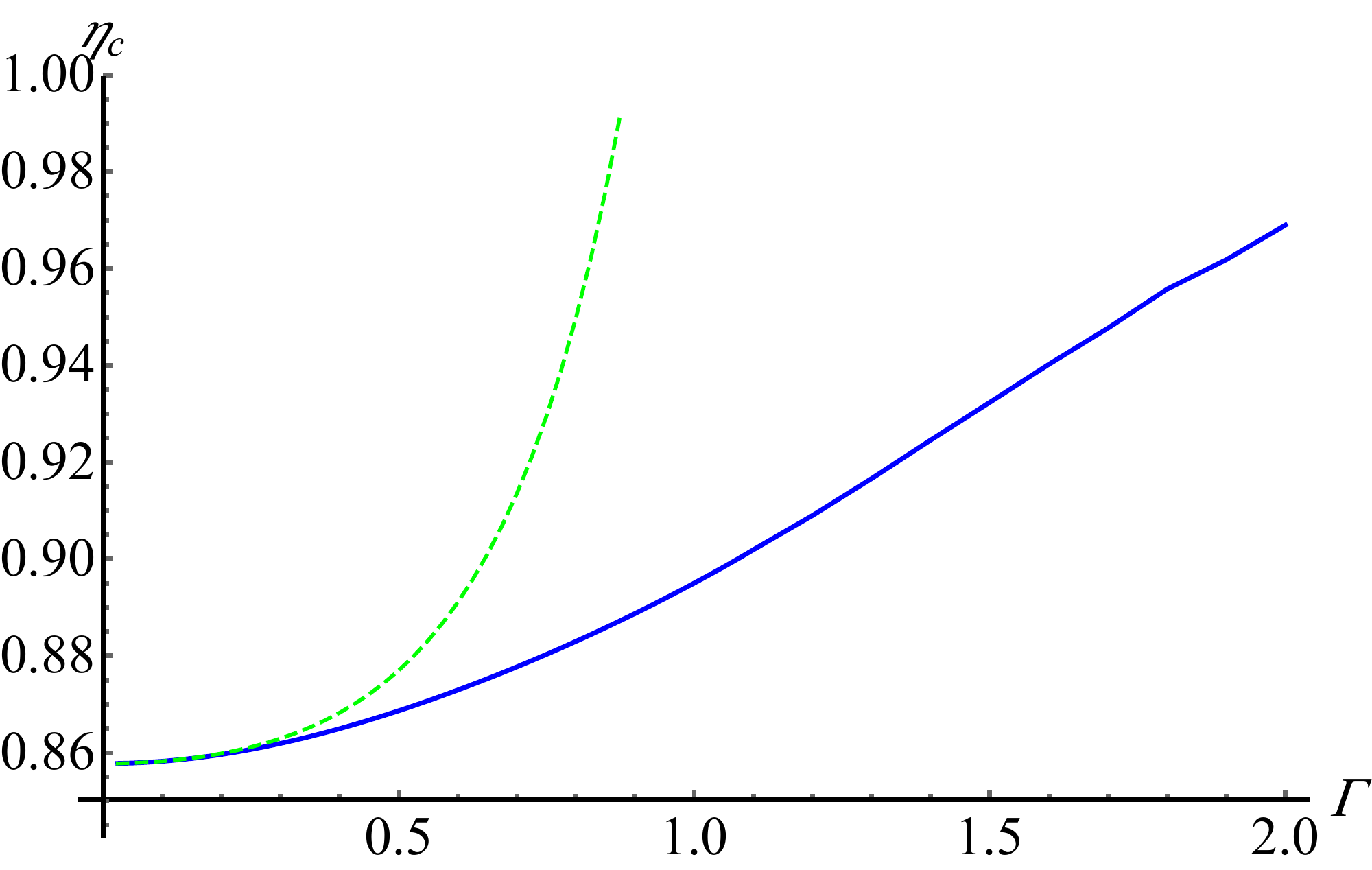} 
\caption{Critical efficiency $\eta_c$ versus $\Gamma$ for the CHSH inequalities for the BSV state. A blue curve represents $\eta_c$ for sign approach and a green dashed curve for normalized Stokes operators. }
\label{effbsv}
\end{figure}

From FIG. \ref{effbsv} we can see that $\eta_c$ for small $\Gamma$ sign and normalized Stokes operators behaves almost identically. However critical efficiency for sign Stokes operators grows slower with $\Gamma$ than for normalized Stokes operators. Also rate of growth of $\eta_c$ from some point starts to decrease with $\Gamma$ in case of sign operator, where it is not the case of normalized Stokes operators. Such change of rate of growth for higher $\Gamma$ should be expected because in case of high number of photons loss of one photon matter less.   
\subsection{CHSH inequality with noise}
\begin{figure}[h!]
\centering
\includegraphics[width=0.49\textwidth]{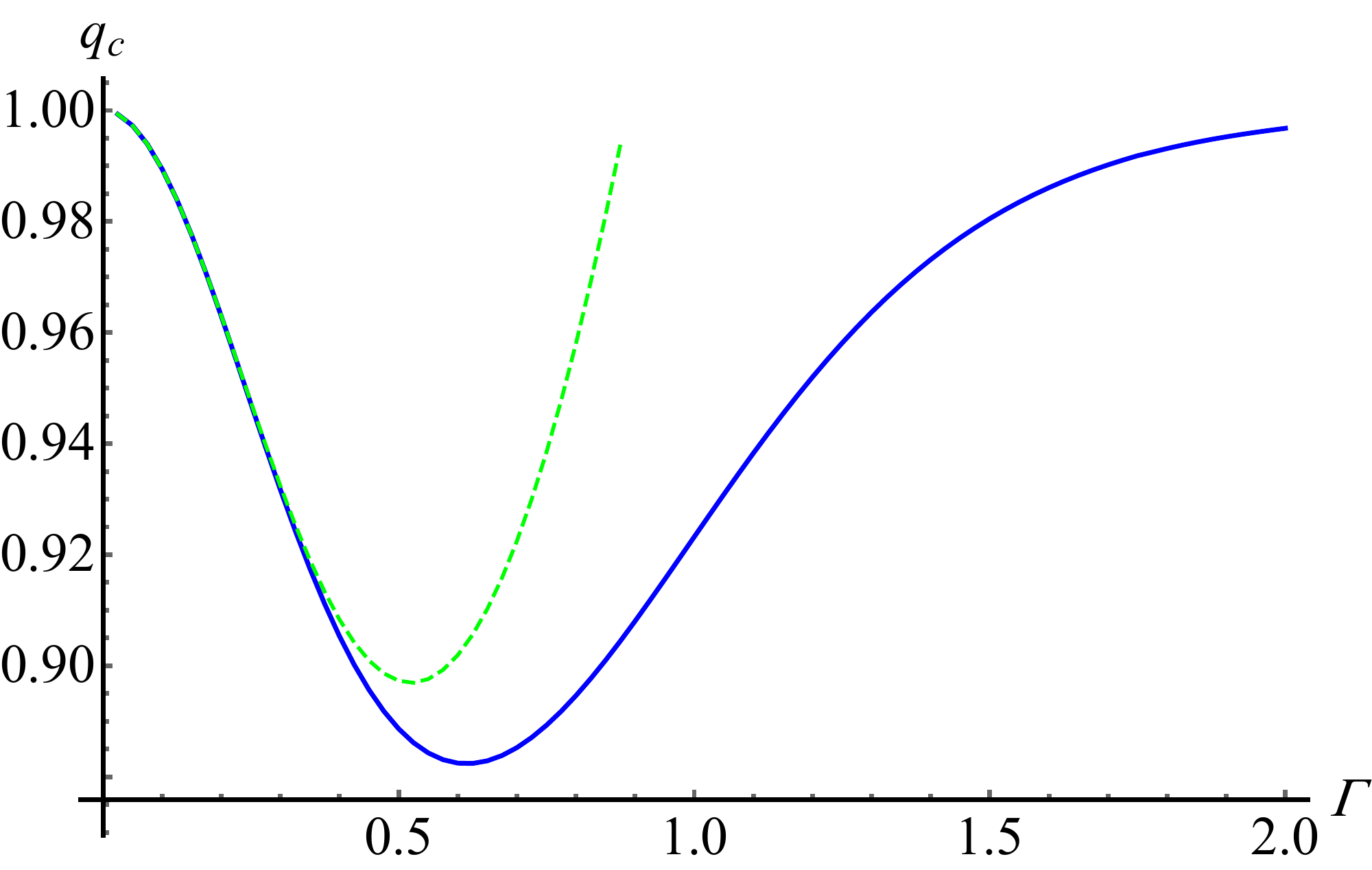} 
\caption{Critical value of $q$ versus $\Gamma$ for the BSV state. A blue curve represents $q_c$ for sign approach and a green dashed curve for normalized stokes operators. Assuming that asymptotic behaviour discussed in \ref{asym} is correct the $q_c$ for the sign Stokes operators goes to 1 in the limit $\Gamma\rightarrow\infty$. }
\label{bsvnoise}
\end{figure}

Another deviation from idealized case in experimental setup can be uncorrelated noise in a state. Let us consider model of white noise with noisy state of the form:
\begin{align}
\begin{split}
\rho'=q\rho_{BSV}+\frac{1-q}{4}(&\ket{\phi^+}\bra{\phi^+}+\ket{\phi^-}\bra{\phi^-}\\
&+\ket{\psi^+}\bra{\psi^+}+\ket{\psi^-}\bra{\psi^-}),
\end{split}
\end{align}
where $\ket{\psi^\pm}$, $\ket{\phi^\pm}$ are bright squeezed vacuum states corresponding to states from Bell basis and $\rho_{BSV}$ is a density matrix of the BSV state. Value $1-q$ determines probability of measuring noise. FIG. \ref{bsvnoise} shows critical value $q_c$ of parameter $q$ for which there is no violation of (\ref{bsvin}) if $q<q_c$ in case of sign and normalized Stokes operators. From FIG. \ref{bsvnoise} one can observe that sign Stokes operators have similar advantage as in case of losses. For small $\Gamma$ the resistance for noise in case of sign Stokes operators is close to this obtained with normalized Stokes operators. However, new operators reach higher maximal resistance and $q_c$ goes asymptotically to 1. 

\subsection{CH inequality}

Going along with idea of sign operators and rate approach to CH inequality \cite{ZUKUBELL} we can construct new CH-like inequality. Let us start with rate operators $\hat R_{+}(i)=\hat\Pi\hat n_i/(\hat n_i+n_{i\perp})\hat\Pi $ used in \cite{ZUKUBELL}. This operator is simply first term of normalized Stokes operator (\ref{STOKES0}). One can observe that it has  eigenvalues in $(1/2,1]$ for states $\ket{n}_i\ket{m}_{i_\bot}$ with $n>m$ and in $[0,1/2)$ for $n< m$. In case of sign Stokes operators we have assigned eigenvalues to those two subspaces of states based on extreme eigenvalues of normalized Stokes operator (corresponding to polarization $i$) for those subspaces. The pattern was to take highest eigenvalue in case $n>m$ and the lowest in the case  $n< m$. We can do the same thing with rate operator with the modification in which we add to second subspace states for $n= m$. As a result we obtain dichotomic observable with eigenvalues: $0$ and $1$. One can observe that Such observable is simply projector onto subspace $n>m$:
\begin{equation}
\hat P(i)=\sum_{n>m}\ket{n}_i\ket{m}_{i_\bot}\bra{n}_i\bra{m}_{i_\bot}.
\end{equation}
From that follows that expectation values of $\hat P^X(i)$ is equal to probability of the observer $X$ to see $n>m$. We shall denote by $\langle \hat P^X(i) \hat P^Y(j)\rangle$ the quantum  joint probability of obtaining the same result $n>m$ by observers $X$ and $Y$ for their respective polarization basis $i$ and $j$.
 Had the probabilities been classical, e.g. Kolmogorovian, and if setting choice at one station cannot directly influence the result on the other one (Einstein's locality) the following Clause-Horne like Bell inequality must hold

\begin{align}
\begin{split}
-1\leq&\Big\langle \hat P^1_+(\theta)\hat P^2_+(\phi)+\hat P^1_+(\theta)\hat P^2_+(\phi')+\hat P^1_+(\theta')\hat P^2_+(\phi)\\
&-\hat P^1_+(\theta')\hat P^2_+(\phi')-\hat P^1_+(\theta)-\hat P^1_+(\phi)\Big\rangle\leq 0. \label{chi}
\end{split}
\end{align}
as classical probabilities of four events $A$, $A'$, $B$ and $B'$ satisfy the Clauser-Horne inequality
\begin{eqnarray}
& -1\leq P(A,B)+P(A, B')&\nonumber \\&+ P(A',B)-P(A',B')-P(A)-P(B)\leq 0.&   
\end{eqnarray}

%TU TRZEBA ZMIENIĆ ZAPIS BO TO NIE MA ZASTOSOWANIA DO KWANTOWYCH WARTOŚCI

\begin{figure}[h!]
\centering
\includegraphics[width=0.49\textwidth]{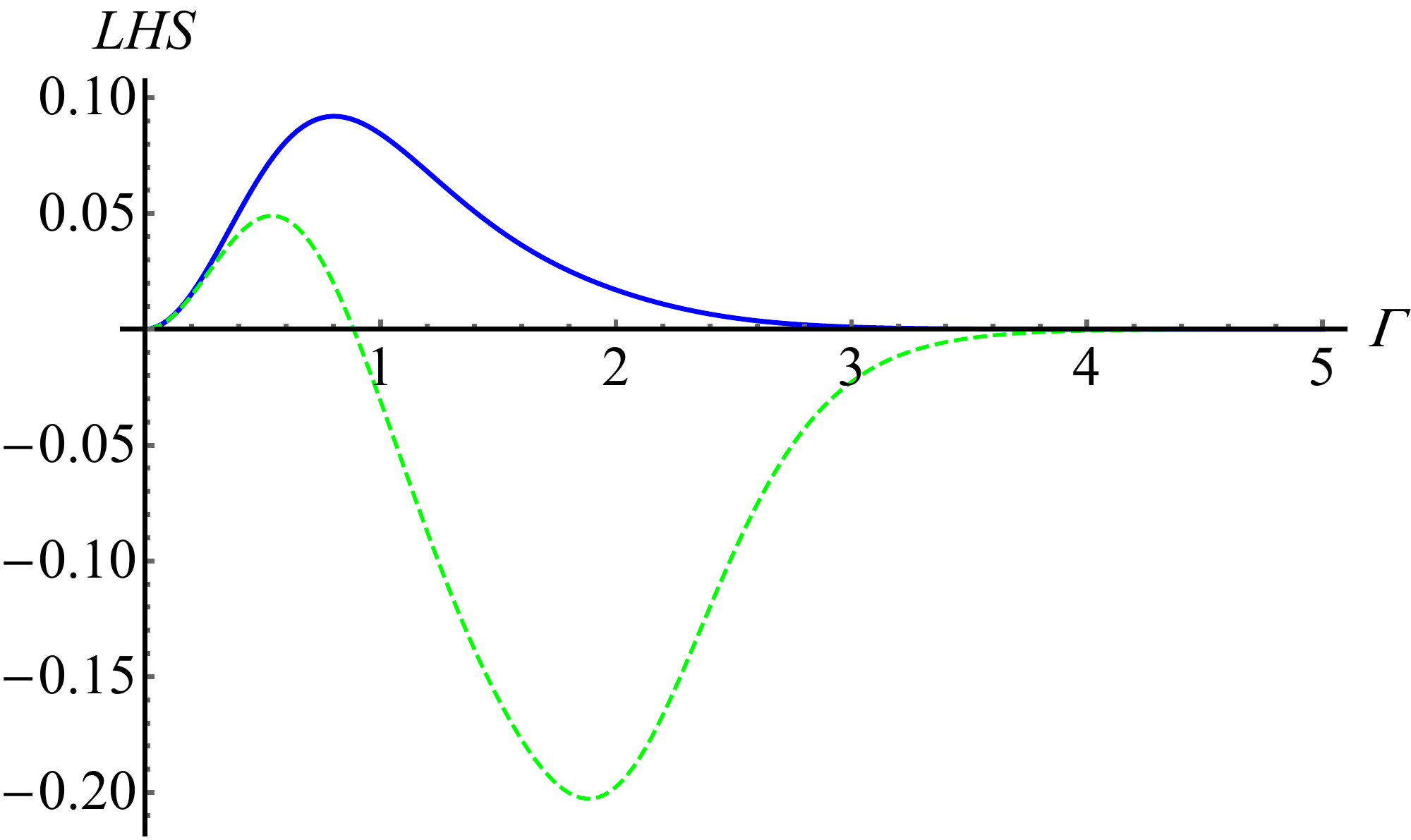} 
\caption{$LHS$ of the CH inequalities for the `sign' approach (blue curve) and rate approach \cite{ZUKUBELL} (green dashed curve) as a function of amplification gain  $\Gamma$ for BSV state (terms up to 50 photons were used in calculations). The CH inequality for the `sign' approach is violated in whole regarded range of $\Gamma$, where the violation in case of the normalized Stokes operators is quickly damped and after that $LHS$ goes asymptotically from bellow to classical bound.  }
\label{chfig}
\end{figure}

FIG. \ref{chfig} shows expectation value of expression  (\ref{chi}) and it's rate counterpart for the same settings as in case of CHSH inequality. The `sign' approach gives  for all $\Gamma$ while the  rate approach gives a violation only for $\Gamma<0.8866$ which is the same case as for CHSH. Note that this CH inequality is not equivalent to CHSH inequality (\ref{bsvin}) (see Appendix \ref{psinV})

\section{Violation of Bell inequalities with sign approach for Bright GHZ}
In order to check if proposed sign Stokes operators can give advantage over normalized Stokes operators not only in one case, let us consider the Bright GHZ state. This state has a following form: 
\begin{align}
\begin{split}
\ket{BGHZ}=\sum_{k=0}^\infty\sum_{m=0}^k& C_{k-m}(\Gamma)C_m(\Gamma)\\
&\times(\hat a_1^\dagger\hat a_2^\dagger\hat a_3^\dagger )^{k-m}(\hat b_1^\dagger \hat b_2^\dagger \hat b_3^\dagger)^m\ket{\Omega}, 
\label{BGHZ0} 
\end{split}
\end{align}
where coefficients $C_Q(\Gamma)$ can be obtained with method presented in \cite{BGHZ} and $\hat a_X^\dagger$, $\hat b_X^\dagger$ are creation operators in two orthogonal polarization modes of beam which goes to observer $X$.

\subsection{Mermin-like inequality}
Let us consider Mermin-like inequality for quantum optical fields \cite{BGHZ}:
\begin{multline}
|\langle S_1^1(\lambda) S_1^2(\lambda) S_1^3(\lambda)- S_1^1(\lambda)S_2^2(\lambda) S_2^3(\lambda)\\- S_2^1(\lambda) S_1^2(\lambda) S_2^3(\lambda)- S_2^1(\lambda) S_2^2(\lambda) S_1^3(\lambda)\rangle_{LHV}|\leq 2,
\label{BELL}
\end{multline}
where $S_i^X(\lambda)$ are local hidden values corresponding to normalized Stokes operators. This inequality generalize Mermin inequality for three qubits \cite{Mermin:inequality}. Derivation of this inequality requires only that local hidden values are bounded by $\pm 1$. Because local hidden values for sign Stokes operators fulfill this requirement, we can change $S_i^X(\lambda)$ to $G_i^X(\lambda)$ and obtain new inequality.However, such inequality is not violated by BGHZ state as in case of normalized Stokes operators. We have to again modify sign Stokes operators:  
\begin{equation}
\hat G_i^X\rightarrow \hat G_i^{X-}=\hat G_i^X-\ket{\Omega_X}\bra{\Omega_X}.
\end{equation}
One can easily write modified local hidden values for such operators as in \ref{sectionchsh} and obtain inequality:
\begin{align}
\begin{split}
\label{BELLNONVAC}
|\langle G_1^{1-}(\lambda)G_1^{2-}&(\lambda)G_1^{3-}(\lambda)\\
- G_1^{1-}(\lambda)G_2^{2-}&(\lambda)G_2^{3-}(\lambda)- G_2^{1-}(\lambda)G_1^{2-}(\lambda)G_2^{3-}(\lambda)\\
&- G_2^{1-}(\lambda)G_2^{2-}(\lambda)G_1^{3-}(\lambda)\rangle_{LHV}|\leq 2.
\end{split}
\end{align}

\begin{figure}[h!]
\centering
\includegraphics[width=0.49\textwidth]{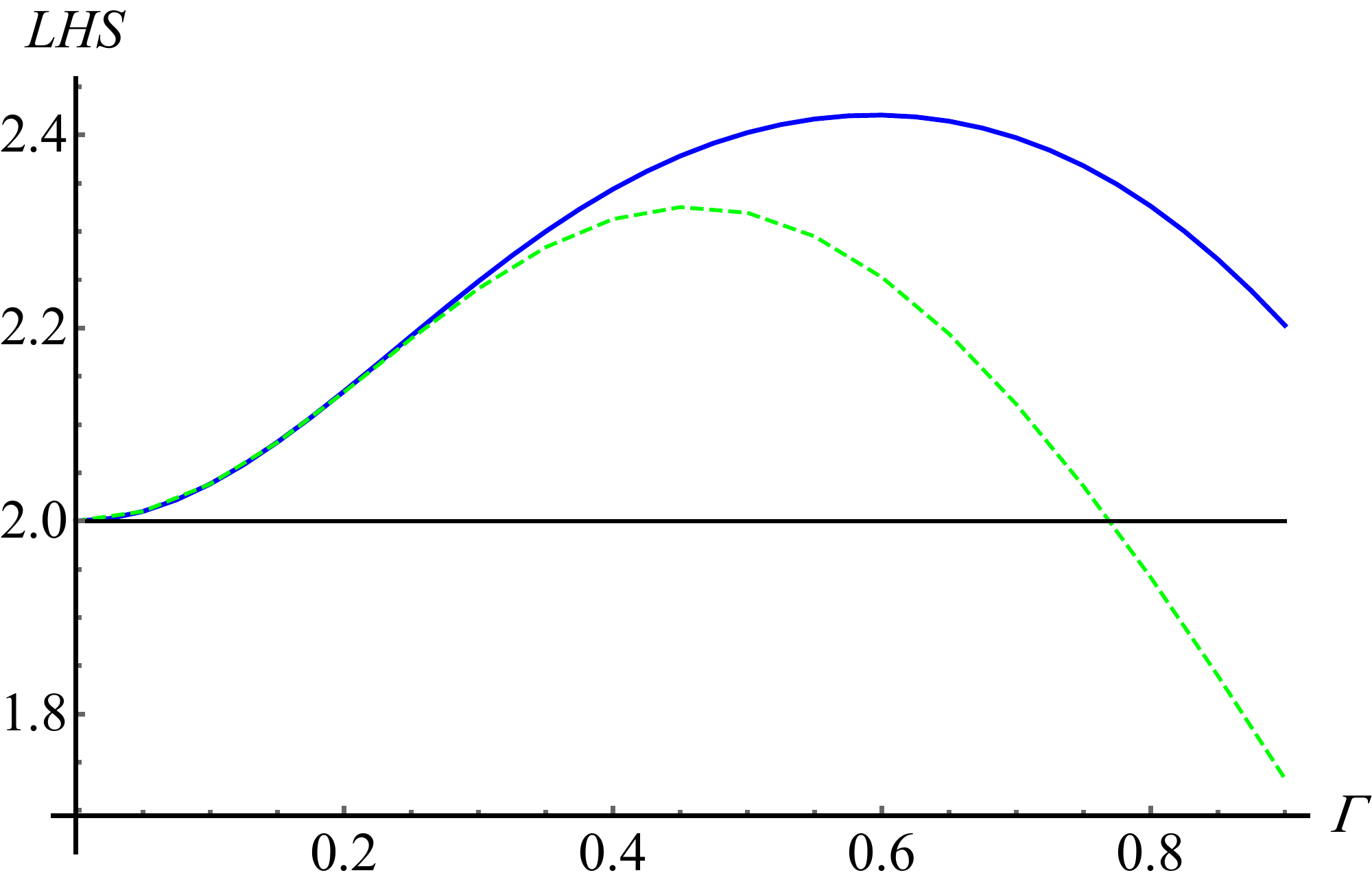} 
\caption{LHS of Bell inequality in function of parameter  $\Gamma$ for BGHZ state}
\label{bellghz}
\end{figure}

FIG. \ref{bellghz} presents LHS of inequality (\ref{BELLNONVAC}) and LHS of analogous inequality for normalized Stokes operators as a function of amplification gain  $\Gamma$. Range of  $\Gamma$ for which inequality is violated by BGHZ state in case of sign Stokes operators  exceeds the range of applicability of the method used in approximating probability amplitudes for BGHZ state. We also stress that this result is more robust than in case of normalized Stokes operators.   

\subsection{Mermin-like inequality with losses}
\begin{figure}[h!]
\centering
\includegraphics[width=0.49\textwidth]{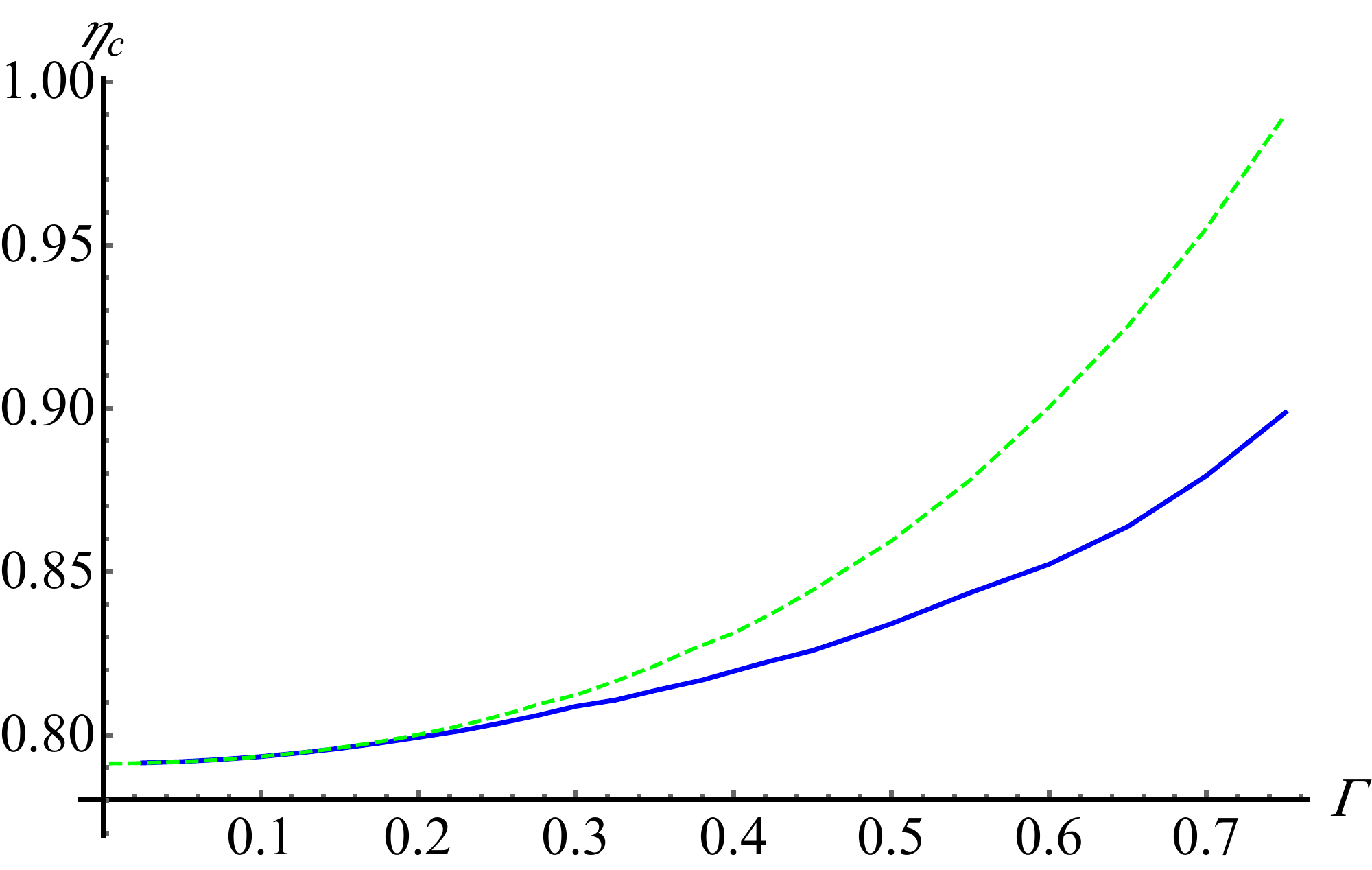} 
\caption{Critical efficiency $\eta_c$ versus $\Gamma$ for the Mermin-like inequalities for the BGHZ state. The blue curve represents $\eta_c$ for sign approach and a green dashed curve for normalized Stokes operators.}
\label{nief}
\end{figure}

We once again consider model of losses on inefficient detectors as in \ref{losses} for inequality (\ref{BELLNONVAC}). On FIG. \ref{nief} critical value of efficiency of detectors $\eta_{c}$ was compered in case of sign and normalized Stokes operators. We can see that for small $\Gamma$ inequalities exhibit similar resistance for inefficient detectors. However, with increasing $\Gamma$  difference between performance of sign operators and normalized operators increases in favor of the first of two.

\section{Conclusions}
%Let us first summarize the results. 
We have proposed new Stokes-like polarization observables for quantum optical fields which have a clear operational meaning. In presented examples the sign Stokes obsevables allow  observation of Bell non-classicality of squeezed-vacuum-type states for pumping powers, for which normalized Stokes observables fail do to so. %Our approach allowed for more robust theoretical prediction of violation of Bell inequalities for the BSV state and the BGHZ state. 

%Conclusion from this research is that 
%The sign Stokes operators may perform better than normalized Stokes operators in multiple cases in exhibiting non-classicality if the broad spectrum of eigenvalues of the latter is not needed from algorithmic standpoint.

%$Sone open questions arises. 

One could be tempted to use sign Stokes operators to derive entanglement indicators not based on Bell inequalities, that is in a form of separability conditions. However those operators do not poses properties which are commonly used in  derivations of bounds for separable states. Thus, this requires a  novel approach. Similar questions arise when one thinks of a steering condition involving sign Stokes operators.

Another question would be if there is a type of state for which normalized Stokes operators allows for violating of some Bell inequality and for which this is impossible using sign Stokes operators.

{Finally, we state that we came across the reference \cite{PhysRevA.104.043323}, when  our manuscript was already written. The paper \cite{PhysRevA.104.043323} introduces the `sign' approach for observables based on particle number measurements in two outputs of measuring device, however in a different context (BEC condensates), and does not relate these with Stokes operators.}

\begin{acknowledgments}
The work is a part of ‘International Centre for
Theory of Quantum Technologies’ project (contract no. 2018/MAB/5), which is carried out
within the International Research Agendas Programme (IRAP) of the Foundation for Polish
Science (FNP) co-financed by the European Union from the funds of the Smart Growth
Operational Programme, axis IV: Increasing the research potential (Measure 4.3).
\end{acknowledgments}
\begin{appendices}
\section{Experimentally friendly properties of sign operators }\label{exper}
\setcounter{equation}{0}

In laboratory one has polarizers and phase shifters and optically active elements to disposal for constructing polarization measurements. Such optical elements perform unitary transformations from which one can obtain transformations $\hat U(H\rightarrow i)$. It can be seen during consideration on how $\hat U(H\rightarrow i)$ acts on creation operators $a^\dagger_H$, $a^\dagger_V$ for $\{D,A\}$, $\{R,L\}$ :
\begin{equation}
\left( \begin{array}{c}
\hat a_{i}^\dagger \\
\hat a_{i_\bot}^\dagger 
\end{array} \right) =
\left( \begin{array}{cc}
\cos{\theta} & \sin{\theta} \\
-\sin{\theta} &\cos{\theta}
\end{array} \right)
\left( \begin{array}{cc}
1 & 0 \\
0 & e^{i\phi}
\end{array} \right)
\left( \begin{array}{c}
\hat a_H^\dagger \\
\hat a_{V}^\dagger 
\end{array} \right),
\end{equation}
where $\theta=\pi/4$ ($\theta=-\pi/4$) and $\phi=0 $ ($\phi=3\pi/2$) for $\{D,A\}$ ($\{R,L\}$).
First matrix is a rotation matrix which transformation can be realized by active element that uses for example Faraday effect to rotate polarization by angle $\theta$.  The second is matrix of a phase shift $\phi$ in second polarization mode.
% It can be also be realized as rotation of polarization analyzer by angle $-\theta$ and relabeling the two resulting polarizations distinguished by analyzer as $\{H, V\}$

In general if we rotate polarization analyzer by some angle  or add some phase shifter in measurment setup we perform unitary transformation on observable changing polarization basis in which we measure. So this transformation is $\hat U(i\rightarrow i')=\hat U_{i'}\hat U_i^\dagger$. Sign Stokes operators under $\hat U(i\rightarrow i')$ transforms in the following way:
\begin{align}
\begin{split}
\hat U(i\rightarrow i')\hat G(i)&\hat U^\dagger(i\rightarrow i')\\
=\mathrm{sign}&(\hat U(i\rightarrow i')\hat U(H\rightarrow i)\\
&(\hat n_H-\hat n_{V})\hat U^\dagger(H\rightarrow i)\hat U^\dagger(i\rightarrow i'))\\
=\mathrm{sign}&(\hat U(H\rightarrow i')(\hat n_H-\hat n_{V})\hat U^\dagger(H\rightarrow i'))\\
=\hat G(i'&).
\end{split}
\end{align} 
In second line we have used fact that in spectral decomposition of $\hat G(i)$ unitary transformation acts only on projectors (preserving their orthogonality) and preserves eigenvalues and thus sign function has no impact on transformation. From this results we see that operators transforms into each other under unitary transformations. Also sign Stokes operators can be realized experimentally analogously to standard Stokes operators.

\section{Violation of Bell inequalities with $\ket{\psi^n_-}$ }\label{psinV}

\begin{figure}[h!]
\centering
\includegraphics[width=0.49\textwidth]{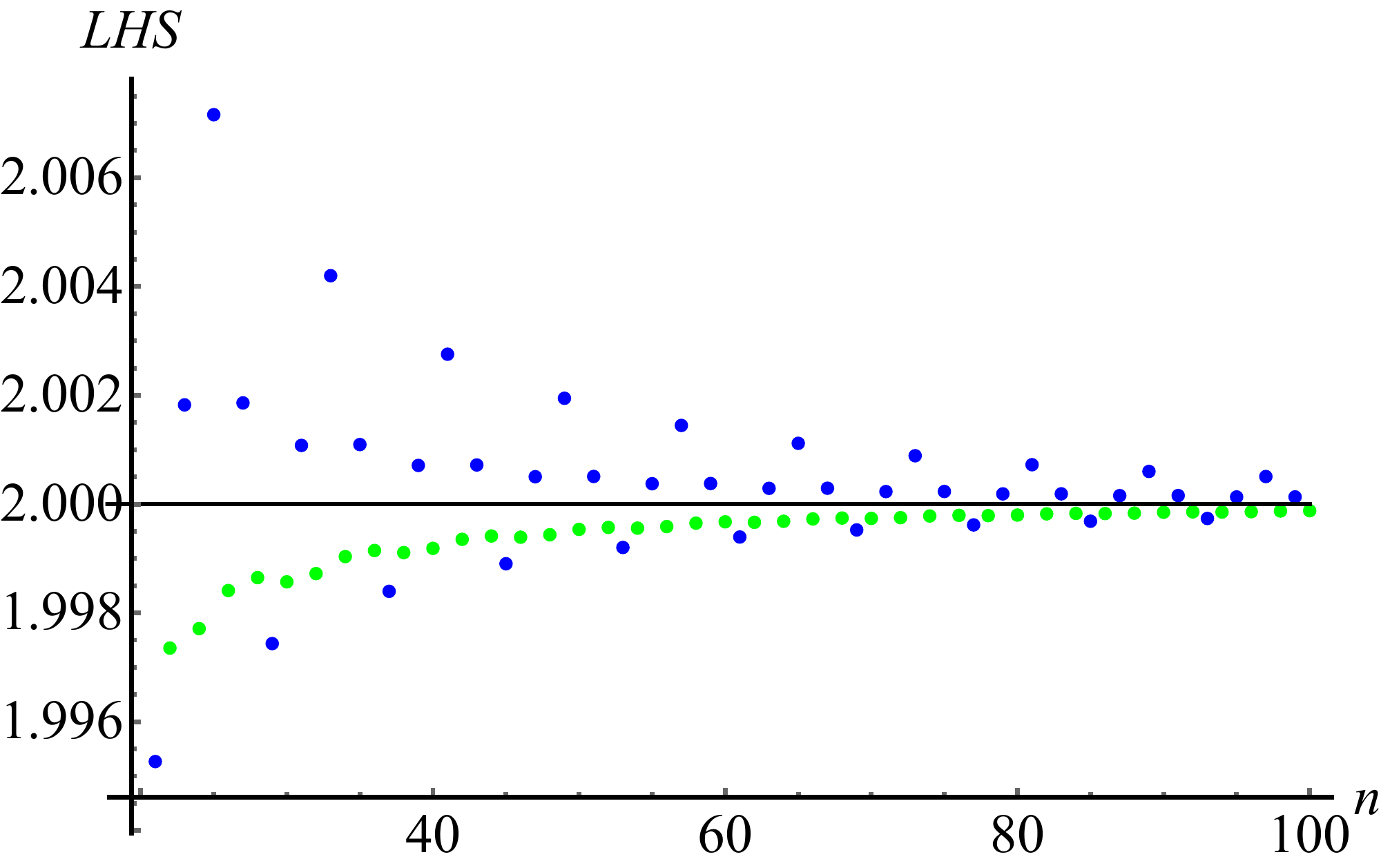} 
\caption{$\langle CHSH\rangle_{\psi^n}$ versus $n$. Blue darker dots depict odd $n$ and green dots stand for even $n$.}
\label{psinZB}
\end{figure}

Let us make more in depth analysis of violation of CHSH inequality (\ref{bsvin}) by $\ket{\psi^n_-}$. Fig. \ref{psinZB} shows violation of (\ref{bsvin}) by $\ket{\psi^n_-}$. We can observe two patterns occurring.The first pattern for odd $n$'s oscillates around bound with decreasing amplitude and period $T=8$. The second pattern for even $n$'s converge to $2$ from bellow with growing $n$. This pattern also have internal structure which is repeatable with $T=8$ (increase, decrease, increase, increase). Note that only $\ket{\psi^n_-}$ with odd $n$ violate CHSH inequality and that in odd pattern for every $n$ which does not violate (\ref{bsvin}) we have $3$ different odd $n$'s for which violation occurs.

Let us make argument why conservation of the pattern for higher $n$'s is expected. Note that $\ket{\psi_-^n}$ can be written in the following form:
\begin{equation}
\ket{\psi_-^n}=\frac{1}{n!\sqrt{n+1}}(\hat a_1^\dagger\hat b_2^\dagger-\hat a_2^\dagger\hat b_1^\dagger)^n\ket{\Omega}.    
\end{equation}
Given some  point in the pattern $\langle CHSH\rangle_{\psi^n}$ to obtain next corresponding point in the pattern we have only to apply operator $(\hat a_1^\dagger\hat b_2^\dagger-\hat a_2^\dagger\hat b_1^\dagger)^T$ to the state $\ket{\psi_-^n}$ and normalize it by the factor $\frac{n!\sqrt{n+1}}{(n+T)!\sqrt{n+T+1}}$. To obtain next $k$-th corresponding point we have to apply this operator $k$ times. Thus, applying such operator has to preserve some internal symmetries. 
There is no reason for existence of $k$ such that it suddenly stops to preserve those symmetries. Therefore, pattern should be continued for any period $N$.

\begin{figure}[ht!]
\centering
\includegraphics[width=0.49\textwidth]{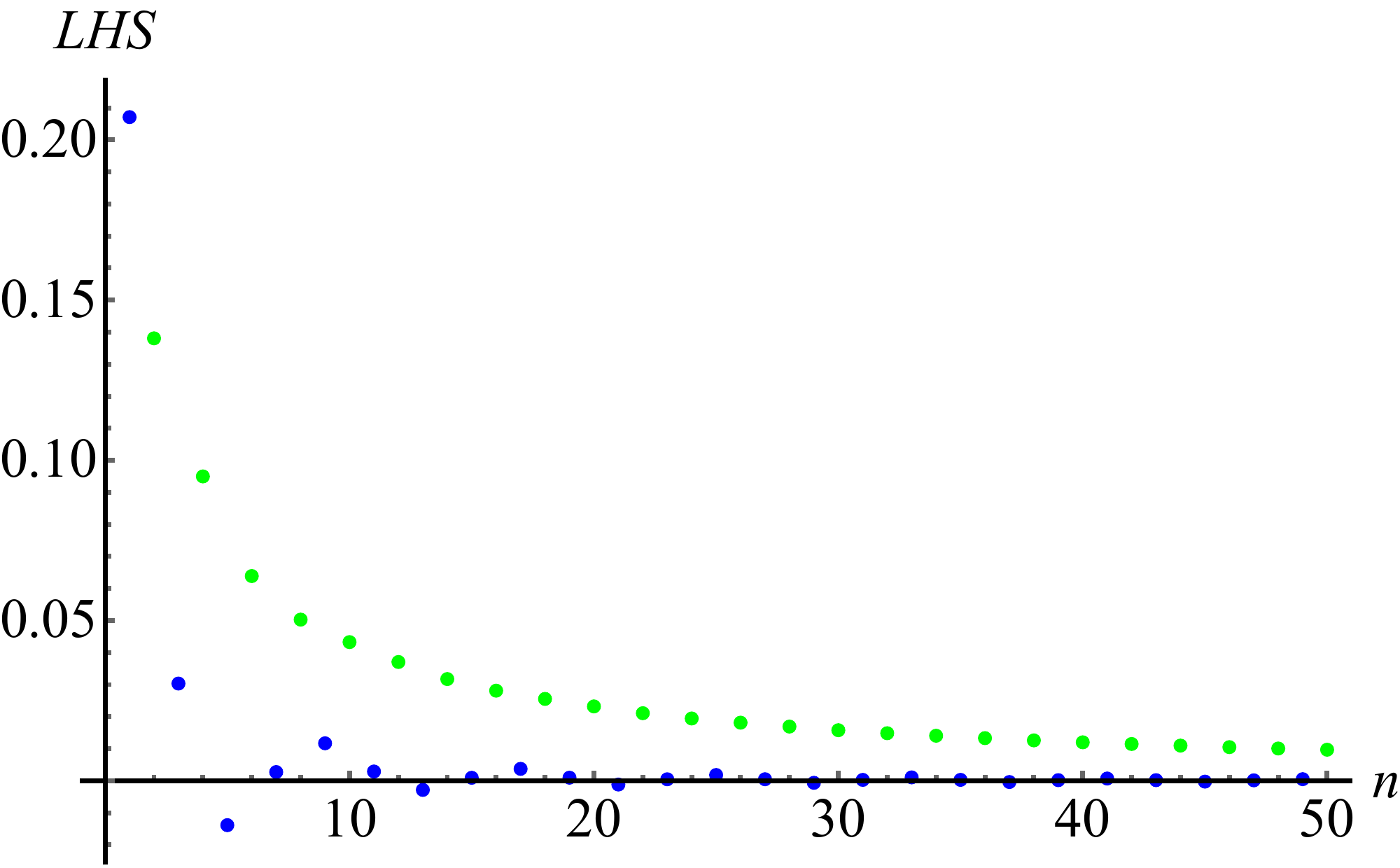} 
\caption{LHS of CH inequality (\ref{chi}) for $\ket{\psi^n_-}$ versus $n$. Blue darker dots depict odd $n$ and green dots stand for even $n$.}
\label{psinch}
\end{figure}

Fig. \ref{psinch} presents LHS of CH inequality (\ref{chi}) for $\ket{\psi^n_-}$. In this case there are also two patterns. The oscillating odd pattern with the same properties and convergent even pattern. However, in this case the even pattern goes to bound from above, and clearly have higher impact on violation of  (\ref{chi}) by the BSV state. This shows That CH inequality (\ref{chi}) is not equivalent to CHSH inequality (\ref{bsvin}).

\end{appendices}

\bibliography{biblio}

\end{document}